\documentclass[english,aps,prl,twocolumn,floats,showpacs]{revtex4}
\usepackage[T1]{fontenc}
\usepackage[latin9]{inputenc}
\usepackage{amsmath}
\usepackage{amssymb}
\usepackage{graphicx}
\usepackage{esint}

\makeatletter

\@ifundefined{textcolor}{}
{%
 \definecolor{BLACK}{gray}{0}
 \definecolor{WHITE}{gray}{1}
 \definecolor{RED}{rgb}{1,0,0}
 \definecolor{GREEN}{rgb}{0,1,0}
 \definecolor{BLUE}{rgb}{0,0,1}
 \definecolor{CYAN}{cmyk}{1,0,0,0}
 \definecolor{MAGENTA}{cmyk}{0,1,0,0}
 \definecolor{YELLOW}{cmyk}{0,0,1,0}
}

\usepackage{babel}
\usepackage{bm}

\makeatother

\begin{document}

\title{Skyrmion Glass in a Disordered Magnetic Film}

\author{E. M. Chudnovsky and D. A. Garanin}

\affiliation{Physics Department, Herbert H. Lehman College and Graduate School, The City University of New York \\
 250 Bedford Park Boulevard West, Bronx, New York 10468-1589, USA}

\date{\today}
\begin{abstract}
We show that high concentration of skyrmions can be achieved in magnetic films with quenched disorder. A 2D system of Heisenberg spins with feromagnetic exchange, $J$, and random magnetic anisotropy of strength $D_R \ll J$ has been studied. Analytical theory  for the dependence of the average skyrmion size on the magnetic field $H$, and for the stability of pinned skyrmions, is complemented by numerical studies of 2D lattices containing up to 40 million spins. At low fields the average size of the skyrmion, $\lambda$, is determined by the average size of Imry-Ma domains. On increasing the field the skyrmions first shrink, with $\lambda \propto D_R/H$, and then collapse at fields distributed around $H_c \propto D_R^{4/3}$. Concentration of the skyrmions goes down with the field as $\exp[-(H/H_c)^{3/2}]$. 
\end{abstract}


\maketitle

The prospect of developing topologically protected data storage has fueled theoretical and experimental research on magnetic skyrmions \cite{Nagaosa2013,Felser-2013,Klaui2016,Fert-Nature2017}. In condensed matter theory they emerge as topologically protected solutions  \cite{SkyrmePRC58,BelPolJETP75,Lectures} of the continuous-field Heisenberg exchange model in two dimensions with the energy $\frac{1}{2}\int d^2 r (\partial {\bf s}/\partial r_\alpha)^2$, where ${\bf s}$ is a three-component spin field of unit length and $r_\alpha = x,y$. An arbitrary spin-field configuration, ${\bf s}({\bf r})$, is characterized by the topological charge, 
\begin{equation}\label{Q}
Q = \frac{1}{8 \pi}\int d^2 r \epsilon_{\alpha\beta} s_a \epsilon_{abc} \frac{\partial s_b}{\partial r_\alpha}\frac{\partial s_c}{\partial r_\beta},
\end{equation}
that takes values $Q = 0, \pm 1, \pm 2$, and so on.  They describe topologically different non-singular mappings of the $(s_x,s_y,s_z)$ unit sphere onto the $(x,y)$ geometrical space. $Q = \pm 1$ correspond to skyrmions and antiskyrmions, while greater $|Q|$ describe multiskyrmion spin-field configurations \cite{GC-condmat}. 

The effect of the weak quenched disorder on the long-range order in systems with continuous-symmetry order parameter has been intensively studied in the past  \cite{ImryMa,CSS,Nattermann-AP2000}. It has been established that the onset of the reversible or glassy behavior depends on the absence or presence of topological defects, as well as on whether the defects are singular or not \cite{PGC-PRL2014}. Non-singular skyrmions in 2D provide weak metastability and narrow hysteresis loop. As we shall see below, the properties of a skyrmion glass formed by quenched randomness in the presence of the magnetic field can be well understood. 

Understanding properties of skyrmions in disordered media can be important for a number of reasons. In a pure exchange model on a lattice, skyrmions collapse due to violation of the scale invariance by the lattice \cite{CCG-PRB2012}. Dipole-dipole interaction stabilizes large skyrmions by forming a skyrmion domain structure, while stability of small skyrmions requires large Dzyaloshinskii-Moriya interaction, anisotropy and/or other than Heisenberg exchange coupling \cite{Bogdanov-Nature2006,Muhlbauer-Science2009,Heinze-Nature2011,Leonov-NatCom2015,Lin-PRB2016,Leonov-NJP2016}. In this paper we show that quenched randomness presents another possibility to stabilize skyrmions. 

The physical mechanism of stabilization of skyrmions by the random anisotropy (RA) is this. In the continuous field approximation the exchange energy of the skyrmion does not depend on its size \cite{Lectures}. Fluctuations of random anisotropy create the potential landscape with the average depth on the spatial scale $R$ that is proportional to the square root of the number of spins, which scales as $R$ in 2D. The Zeeman energy of the skyrmion of size $R$, with the magnetization opposite to the field, scales as $R^2$. This provides the minimum of the energy on the size of the skyrmion that goes down on increasing the field. The discreteness of the crystal lattice provides the dependence of the exchange energy on the skyrmion size. When this effect is added, skyrmions collapse after they reach a minimum critical size on increasing the field. This picture is confirmed by analytical and numerical studies that are in excellent agreement with each other. 

We investigate static properties of a glassy skyrmion phase in a 2D magnetic film with random magnetic anisotropy. The effect of the quenched disorder on the mobility of individual skyrmions has been studied in the past in connection with their current-induced motion \cite{Fert-NatNano2013,Reichhardt-PRL2015}. Our focus is on the field dependence of the concentration of skyrmions, their stability and average size in a centrosymmetric system \cite{Zhang2016}. The model consists of Heisenberg spins subjected to ferromagnetic exchange, the RA, and the external magnetic field. When the anisotropy is stronger than dipole-dipole and Dzyaloshinskii-Moriya interactions, the latter can be neglected. Without taking skyrmions into account this model has been studied in the past \cite{CSS}. Here we show that scaling arguments modified by the presence of skyrmions allow one to describe properties of a skyrmion glass. We begin with working out a theoretical framework for pinning of skyrmions by the RA and then present numerical data on spin lattices containing up to $4 \times 10^7$ spins. Properties of the skyrmion glass that we observe in numerical experiments agree with our analytical results. 

We consider a constant-length spin field ${\bf S}({\bf r})$ with the energy
\begin{equation}\label{Hamiltonian}
{\cal{H}} = \int d^2r\left[\frac{\alpha}{2}(\partial_{\mu}{\bf S}) \cdot (\partial_{\mu}{\bf S}) - \frac{\beta_R}{2}({\bf n}\cdot{\bf S})^2 - {\bf H}\cdot{\bf S}\right],
\end{equation}
where ${\bf n}({\bf r})$ is a unit vector of the RA and ${\bf H}$ is the magnetic field applied in the $z$-direction. Parameters $S$, $\alpha$, and $\beta_R$ are related to the parameters of the lattice model with the energy 
\begin{equation}
{\cal H}_l=-\frac{J}{2}\sum_{ij}{\bf s}_{i}\cdot{\bf s}_{j}-\frac{D_R}{2}\sum_{i}({\bf n}_{i}\cdot{\bf s}_{i})^2-{\bf H}\cdot\sum_{i}{\bf s}_{i}
\end{equation}
via  $S = s/a^2$, $\alpha = Ja^4$, and $\beta_R = 2D_Ra^2$, where $a$ is the lattice spacing. At $H = 0$ the system breaks into Imry-Ma domains \cite{ImryMa} which corresponds to a finite ferromagnetic correlation length \cite{CSS,Lectures}, $R_f /a \propto J/D_R$. The argument goes like this. If a significant rotation of the magnetization were to occur on a scale $R$, the density of the exchange energy would be of order $J/R^2$, while fluctuations of the RA energy density, that provide the coherent anisotropy, would be of order $-(D_R/R^2) [(R/a)^2]^{1/2} = -D_R/(a R)$. Minimization of the sum of the two energies on $R$ yields the above scaling of $R_f$. 

This argument should be modified in the presence of topological defects. Such defects in the isotropic 2D exchange model are Belavin-Polyakov skyrmions. In a continuous spin-field approximation, their energy, due to the scale invariance, does not depend on the skyrmion size $\lambda$. The RA and the Zeeman energies, being small compared to the exchange, cannot significantly affect the shape of the skyrmion. However, since they break the scale invariance, these energies must depend on $\lambda$. At $H = 0$ the only scale generated by the RA is the average size of the Imry-Ma domain $R_f$, which should determine the $H = 0$ scale of $\lambda$. The $z$-component of the spin-field of the skyrmion located at the origin in the background of spins looking up is given by \cite{Lectures} $S_z/S = (x^2 + y^2 - 4\lambda^2)/(x^2 + y^2 + 4\lambda^2)$. In the presence of the field ${\bf H} = H\hat{z}$ it generates Zeeman energy 
$E_Z = -\int d^2r H(S_z - S)  = 16\pi S H \lambda^2 \ln({L}/{2\lambda})$, where we assumed $2\lambda \ll L$. Writing $E_Z$ as $HS \pi R^2(\lambda)$ one can define the magnetic radius of the skyrmion, $R(\lambda) = 4\lambda \ln^{1/2}({L}/{2\lambda})$. In the presence of the RA the system size $L$ should be replaced by the length of order $R_f$. In what follows we will approximate the magnetic size of the skyrmion and its Zeeman energy by 
\begin{equation}\label{REz}
R(\lambda) = 4\lambda \sqrt{l}, \quad E_Z = 16\pi l H S \lambda^2 = 16\pi l s H \left({\lambda}/{a}\right)^2,
\end{equation}
 where $l$ is a logarithmic factor that we shall treat as a constant.

Statistical fluctuations of the RA result in a coherent magnetic anisotropy on the scale $R(\lambda)$. Its strength is proportional to the square root of the number of spins on that scale, which is linear on $R(\lambda)$. The corresponding pinning energy is given by
\begin{equation}\label{A}
E_A = - \frac{\beta_R}{2}\sqrt{\left\langle\left[\int d^2 r ({\bf n}\cdot{\bf S})^2 \right]^2\right\rangle - \left\langle\int d^2 r ({\bf n}\cdot{\bf S})^2\right\rangle^2}
\end{equation}
In a continuous spin-field model, instead of working with a fixed-length random field ${\bf n}$, it is easier to perform calculation with a Gaussian vector field ${\bf h}({\bf r}) \propto {\bf n}({\bf r})$ of the average strength $h = \sqrt{\beta_R/2}$, having $\langle h_{\alpha}({\bf r}')h_{\beta}({\bf r}'')\rangle = \frac{1}{3}h^2\delta_{\alpha\beta}\Gamma(|{\bf r}' - {\bf r}'|/a)$ with $\Gamma(0) = 1$ and $\Gamma_{a \rightarrow 0} \rightarrow \delta[({\bf r}' - {\bf r}'')/a]$ defined as a nascent delta function. Writing $\langle h'_{\alpha}h'_{\beta}h''_{\delta}h''_{\gamma}\rangle = \langle h'_{\alpha}h'_{\beta}\rangle\langle h''_{\delta}h''_{\gamma}\rangle +\langle h'_{\alpha}h''_{\delta}\rangle\langle h'_{\beta}h''_{\gamma}\rangle + \langle h'_{\alpha}h''_{\gamma}\rangle\langle h'_{\beta}h''_{\delta}\rangle$, where ${\bf h}' = {\bf h}({\bf r}'),  {\bf h}'' = {\bf h}({\bf r}'')$, one obtains from Eq.\ (\ref{A}) $E_A = -\frac{1}{3}\beta_RaS^2\sqrt{\int d^2r} $ for the average fluctuation of the RA in the area of size  $\int d^2r$. We use the fact that ${\bf S}({\bf r})$ in a skyrmion changes on a scale $R(\lambda)$ that is much greater than the scale, $a$, of the RA change, which decouples the averaging over ${\bf h}({\bf r})$ from ${\bf S}({\bf r})$. Replacing $\int d^2r$ with $\pi R^2({\lambda})$, we have
\begin{equation}
E_A = -\frac{4}{3}\sqrt{\pi l}\beta_RS^2a\lambda = -\frac{8}{3}\sqrt{\pi l }D_Rs^2\frac{\lambda}{a},
\end{equation}
in accordance with our expectation that fluctuation of the RA energy is linear on the skyrmion size.

The discreteness of the lattice breaks the scale invariance of the exchange interaction. The corresponding energy, $E_l = -(2\pi Js^2/3)(a/\lambda)^2$, has been computed in Ref.\ \onlinecite{CCG-PRB2012}. The total energy to be minimized is 
\begin{equation}\label{competing}
E = 16 \pi l H \lambda^2  - \frac{8}{3}\sqrt{\pi l}D_R \lambda - \frac{2\pi}{3\lambda^2}. 
\end{equation}
In this formula and below we have switched to dimensionless variables by expressing the Zeeman factor $sH$ and the anisotropy constant $D_R$ in the units of the exchange energy $Js^2$ and expressing $\lambda$ in the units of the lattice spacing $a$. Eq.\ (\ref{competing})  describes three competing effects: The positive Zeeman energy that wants to shrink the skyrmion because its magnetic moment is opposite to the field, the negative pinning energy due to the fluctuations of random anisotropy that wants to expand the skyrmion, and the negative lattice energy that wants to collapse the skyrmion. 

\begin{figure}
\includegraphics[width=88mm]{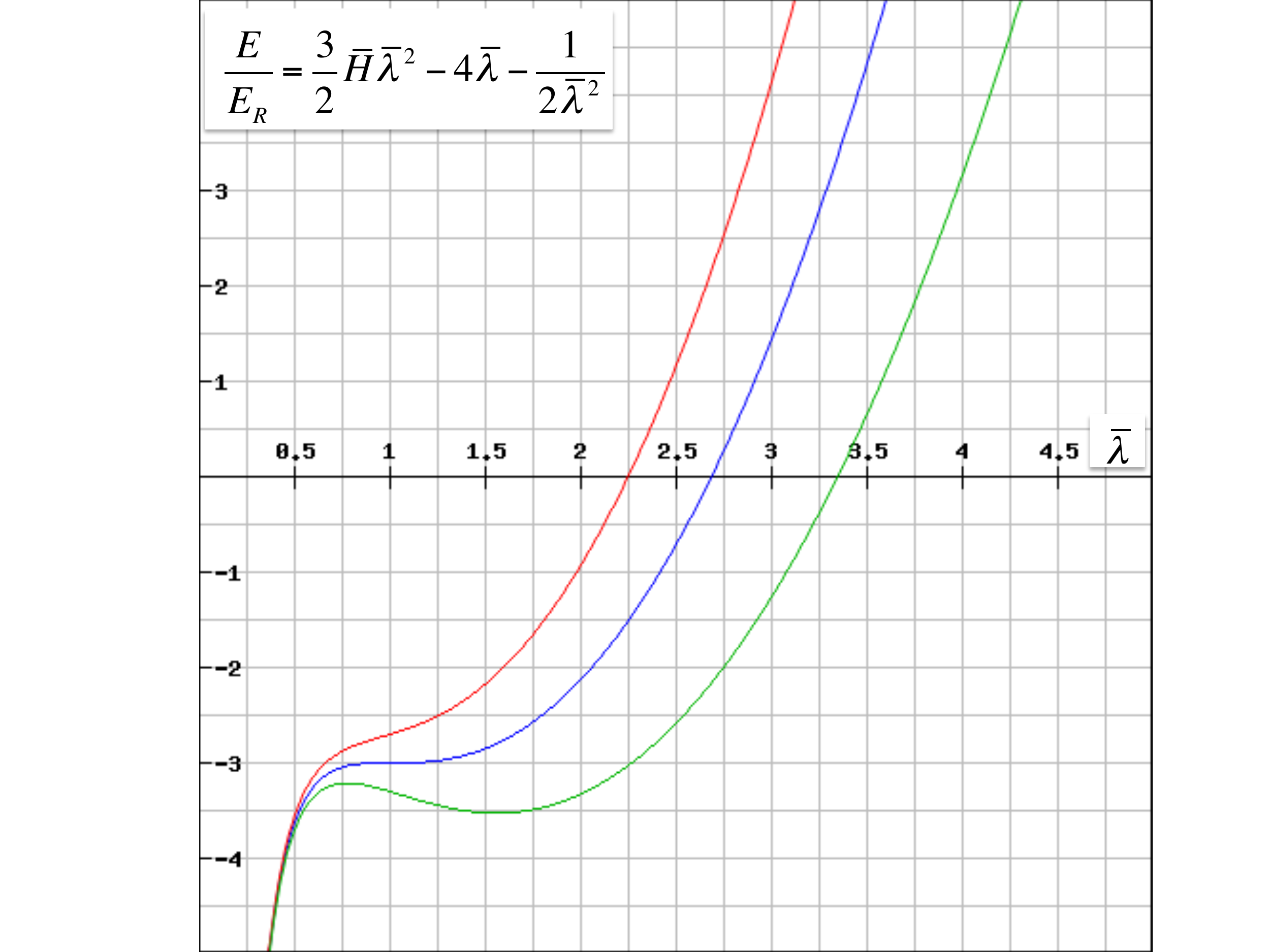}
\caption{Color online: Dependence of energy (\ref{E-tot}) on the skyrmion size at $H < H_c$ (green), $H = H_c$ (blue), and $H > H_c$ (red).}
\label{E}
\end{figure}
Introducing
\begin{equation}\label{critical}
H_c = \frac{1}{32l^{1/3}}\left(\frac{2}{\pi}\right)^{2/3} D_R^{4/3}, \qquad \lambda_c = \left(\frac{4\pi}{l}\right)^{1/6}\frac{1}{D_R^{1/3}}
\end{equation}
it is convenient to write the total energy in the form
\begin{equation}\label{E-tot}
E = E_R \left(\frac{3}{2}\bar{H}{\bar{\lambda}}^2 -4\bar{\lambda} - \frac{1}{2{\bar{\lambda}}^2}\right)
\end{equation}
where $\bar{H} = H/H_c$, $\bar{\lambda}=\lambda/\lambda_c$, and $E_R =  \left(16\pi^2lD_R^2/27\right)^{1/3}$.  It has a maximum and a minimum at $\bar{H} < 1$, and an inflection point at $\bar{H}=1$ when $\bar{\lambda}=1$, see Fig.\ \ref{E}. At $\bar{H} \ll 1$ the minimum is dominated by the first two terms in Eq.\  (\ref{E-tot}), leading to $\bar{\lambda} = {4}/({3\bar{H}})$ and $\lambda = (12\sqrt{\pi l})^{-1}({D_R}/{H})$. Note that the skyrmion size given by this equation is bounded by the length of order $R_f$ at small $H$. The energy minimum disappears at $H = H_c$, which corresponds to skyrmion collapse at $H > H_c$ and $\lambda < \lambda_c$. 

The above formulas help to understand the behavior of skyrmions due to their pinning by the RA. Fluctuations of the pinning energy must lead to the distribution of skyrmion sizes as well as to the distribution of $H_c$ and $\lambda_c$ around the values given by Eq.\ (\ref{critical}). Random factor $p$ in the pinning energy makes the critical field proportional to $p^{4/3}$. Gaussian distribution over $p$ then leads to the $\exp(-p^2) = \exp[-(H/H_c)^{3/2}]$ distribution over critical fields. This, as well as the $D_R/H$ scaling of the average skyrmion size at intermediate fields, is supported by the numerical experiment on spin lattices. In comparing our theory with numerical results we are after functional dependences of physical quantities on the strength of quenched disorder. As we shall see below, the agreement between theory and numerical experiment is rather remarkable, thus confirming our picture with a high degree of confidence.

\begin{figure}
\vspace{5mm}
\includegraphics[width=88mm]{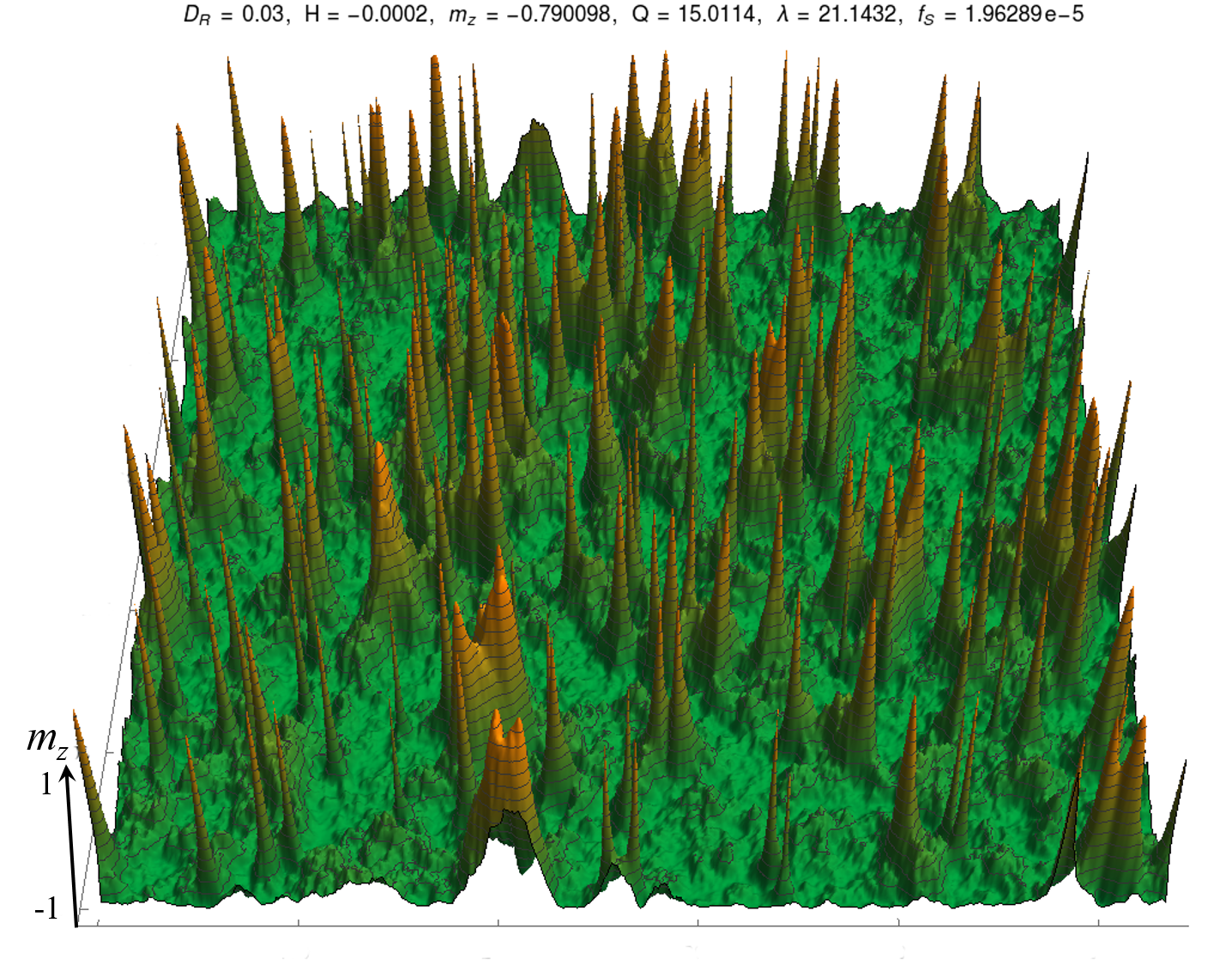}
\includegraphics[width=88mm]{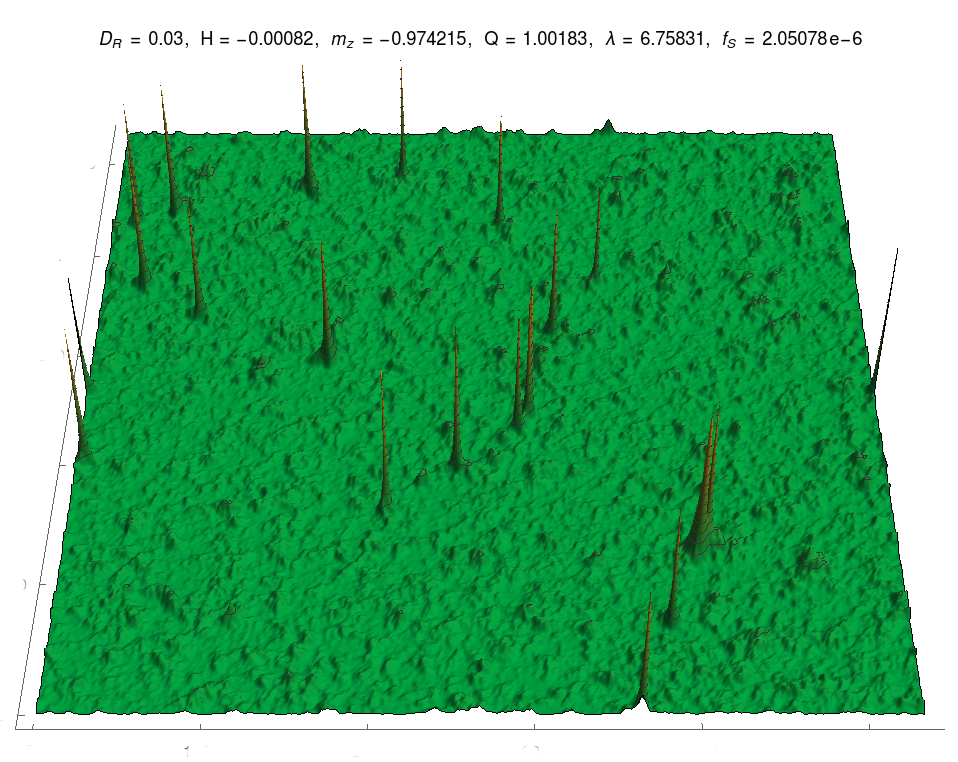}
\vspace{2mm}
\caption{Color online: Skyrmion forest in the 2D RA magnet with $D_R/J = 0.03$ and $3200^2$ spins, obtained by applying a negative field $H$, in small steps, to a random initial spin state at $H=0$. Upper panel: Weak field, $ |H|/J=0.0002$, many large skyrmions, $\langle\lambda\rangle \sim 21$. Lower panel: Stronger field,  $|H|/J=0.00082$, most of the skyrmions have already collapsed, the survivors shrank, $\langle\lambda\rangle \sim 6.8$. Images show the $z$-component of the magnetization.}
\label{forest}
\end{figure}
We minimize the energy of the spin system numerically in the presence of the exchange, the RA which direction is chosen randomly at each lattice site, and the external field on 2D lattices of size $L \gg R_f$. Our numerical method combines sequential rotations of spins towards the direction of the local effective field, ${\bf H}_{i,{\rm eff}}=\sum_{i}J_{ij}{\bf s}_{j}+{\bf h}_{j}+{\bf H}$,
with energy-conserving spin flips: ${\bf s}_{i}\rightarrow2({\bf s}_{i}\cdot{\bf H}_{i,{\rm eff}}){\bf H}_{i,{\rm eff}}/H_{i,{\rm eff}}^{2}-{\bf s}_{i}$, applied with probabilities $\gamma$ and $1-\gamma$ respectively; $\gamma$ playing the role of the relaxation
constant. High efficiency of this method for glassy systems under the condition $\gamma \ll1$ has been demonstrated by us in the past, see, e.g., Ref. \onlinecite{PGC-PRL2014}. The software used was Wolfram Mathematica that allows compilation (including usage of an external C compiler that doubles the speed) and parallelization. The main operating computer was a 20-core Dell Precision T7610 Workstation. The largest-scale computation, that lasted a few days, has been done on a square lattice containing $6400 \times 6400$ spins. 
\begin{figure}
\vspace{5mm}
\includegraphics[width=88mm]{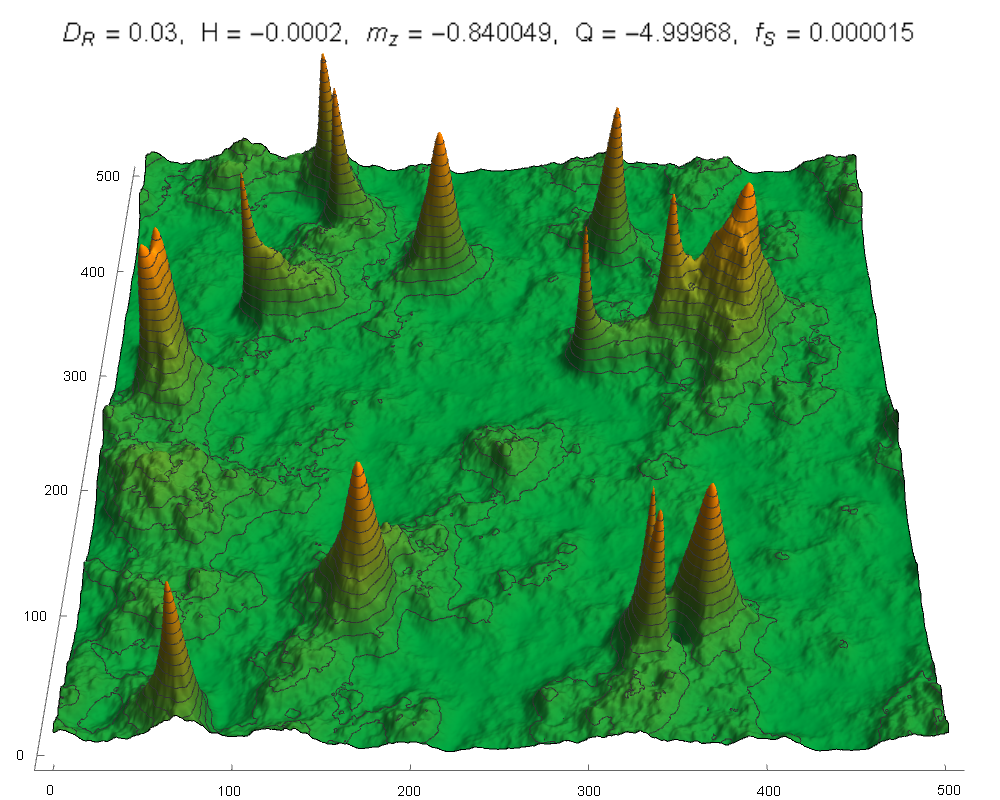}
\includegraphics[width=75mm]{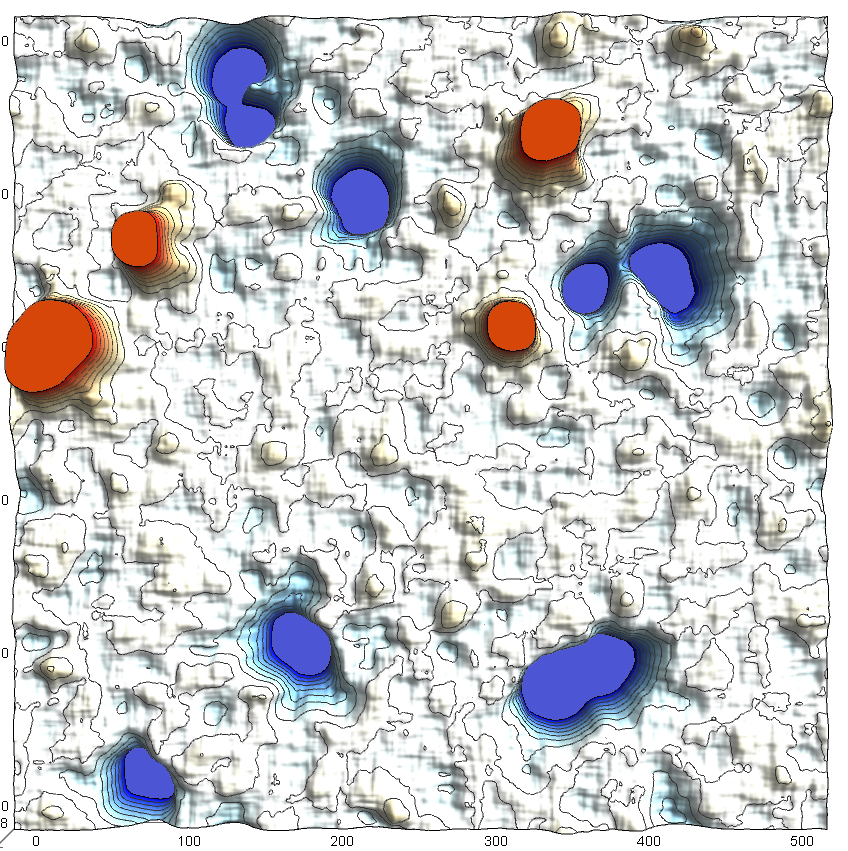}
\vspace{2mm}
\caption{Color online: Correlation between magnetization maxima (upper panel) and density of the topological charge (lower panel). Red (positive) corresponds to skyrmions, while blue (negative) corresponds to antiskyrmions. The charge density beyond the plotting range is shown by uniform color. Adding charges of four skyrmions and nine antiskyrmions (that include two biskyrmions), one arrives to $Q = -5$ in accordance with the displayed value of $Q = -4.99968$ obtained numerically by integration over the whole area.}
\label{topology}
\end{figure}
 
 \begin{figure}
\includegraphics[width=88mm]{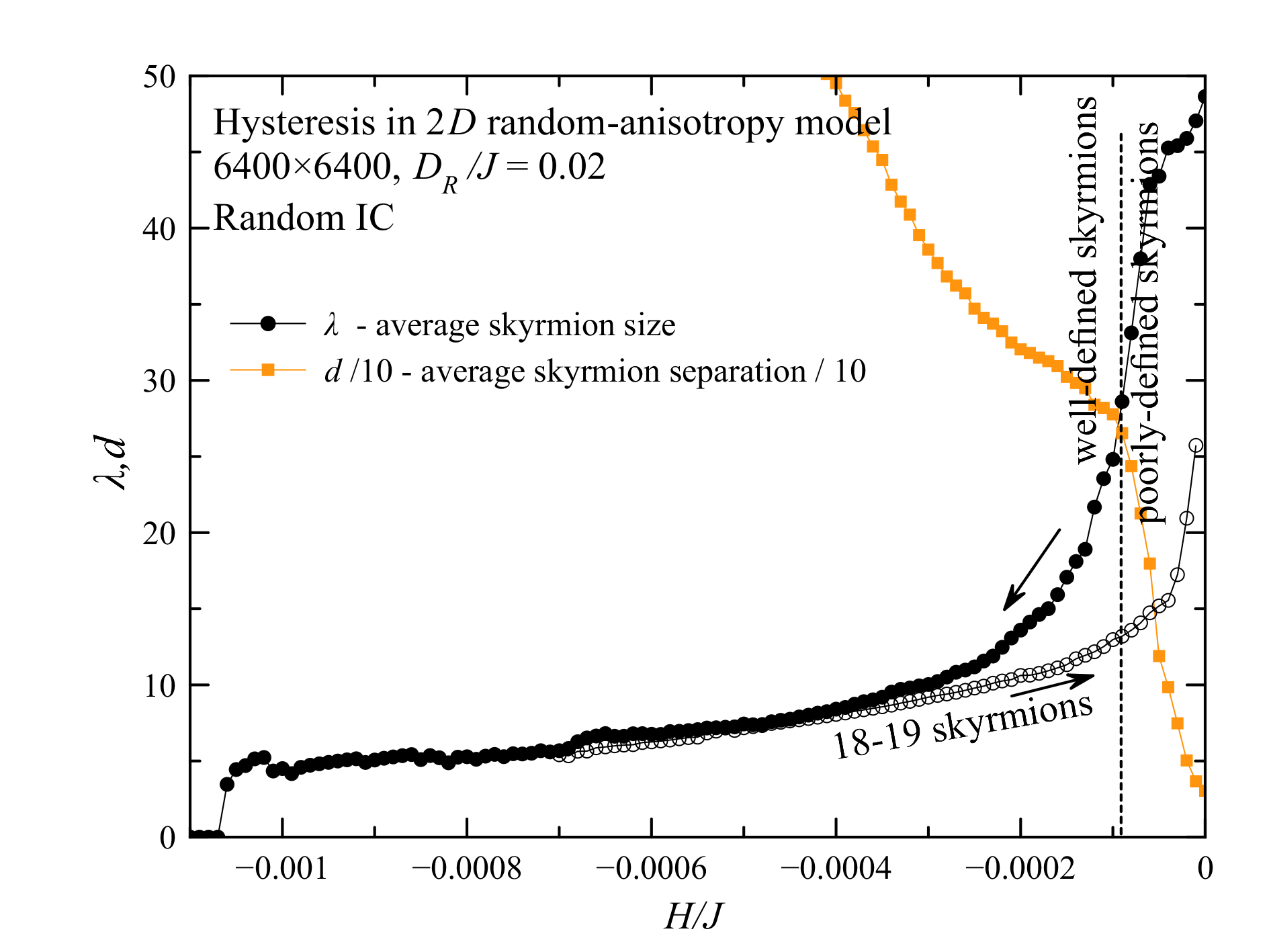}
\caption{Color online: Evolution of the average skyrmion size and the average distance between the skyrmions on changing the field.}
\label{size-H}
\end{figure}
We begin at $H = 0$ and apply the field in the negative direction, in small steps, each time minimizing the energy of the system. Topological charge, $Q$, computed on the lattice by using discretized form of Eq.\ (\ref{Q}), is very close to the integer. It is typically small and is due to the statistical difference between the numbers of skyrmions and antiskyrmions. In counting them we do not distinguished between the two. The total number of skyrmions and antiskyrmions is found by looking at the maxima of the $z$-component of the magnetization, $m_z$. At $H = 0$ skyrmions are comparable in size to the Imry-Ma domains and are difficult to identify. When the field is applied, the well-separated compact skyrmions with the magnetization opposite to the background emerge, see Fig.\ \ref{forest}. Most of the observed topological defects are simple skyrmions and antiskyrmions with $Q = \pm 1$, while a small fraction are biskyrmions. At small fields concentration of skyrmions is high. As the absolute value of the field increases, skyrmions shrink and begin to collapse. Correlation between the magnetization maxima and the density of the topological charge, confirming that we are dealing with skyrmions, is illustrated in Fig.\ \ref{topology}.

The average skyrmion size in a weak field is independent of $H$, see Fig.\ \ref{size-H}, in accordance with our expectation that it is determined by $R_f \sim (J/D_R)a$ at $H = 0$. As the field continues to increase in the negative $z$-direction, the skyrmions go down in size and become less abundant. The average skyrmion separation $d$ shown in Fig.\ \ref{size-H} is defined as $1/\sqrt{f_S}$, where $f_S$ is concentration of skyrmions, that is, their number per spin. On decreasing the field back to zero no new skyrmions are formed but the existing skyrmions grow in size to adjust to the pinning potential. The sizes of skyrmions in the numerically obtained skyrmion forest were evaluated by computing second derivatives of $m_z$ at the maxima and comparing it with the known profile of the Belavin-Polyakov skyrmion. 

\begin{figure}
\includegraphics[width=88mm]{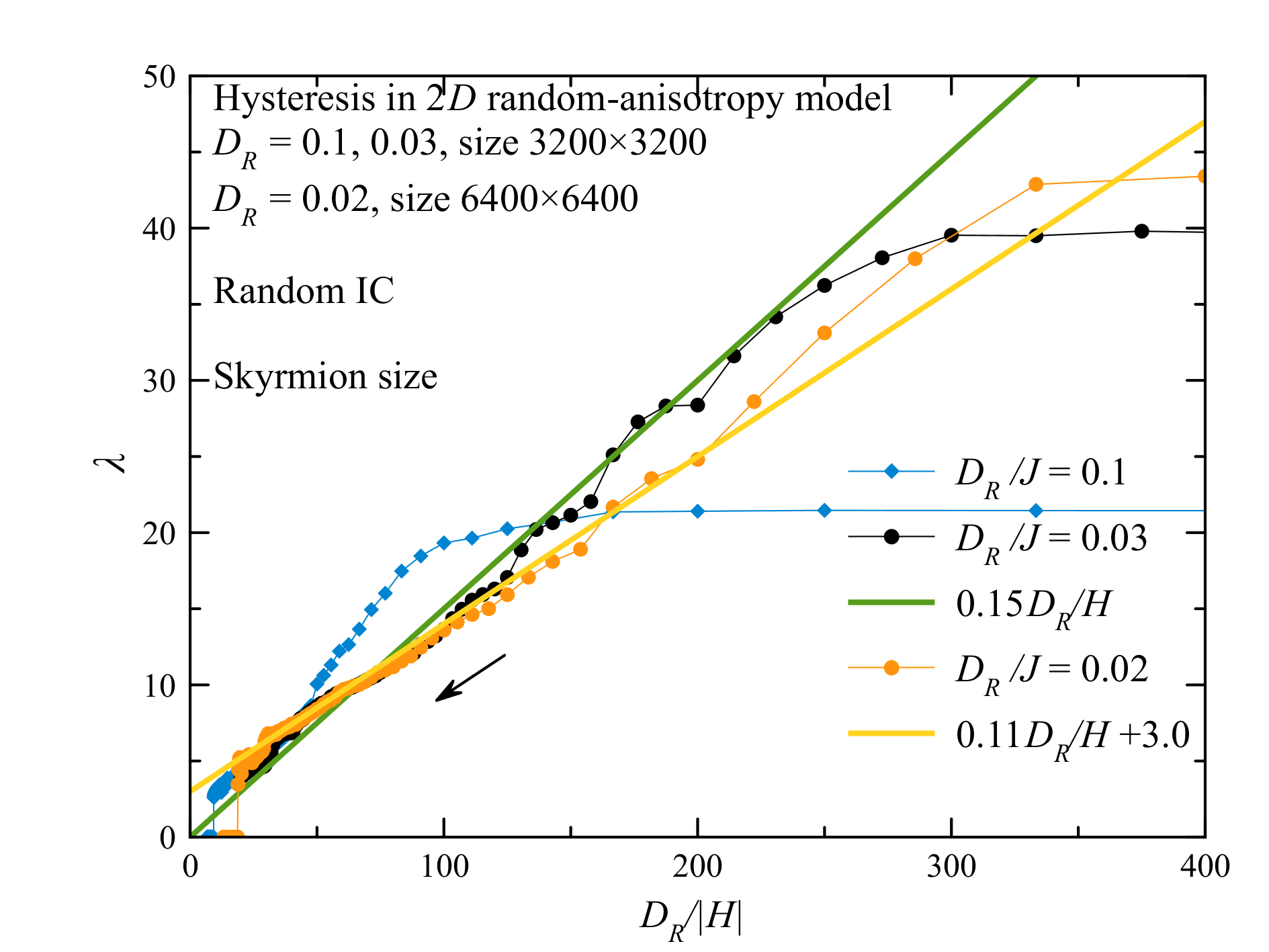}
\caption{Color online: Dependence of the average skyrmion size on $D_R/|H|$.}
\label{size-1/H}
\end{figure}

The dependence of the average skyrmion size on $D_R/|H|$ on increasing the field is shown in Fig.\ \ref{size-1/H}. In accordance with analytical results, it is close to linear in the intermediate field range, saturates at small fields, and shows collapse of the smallest skyrmions at high fields. Linear part of the graph allows one to extract the parameter $l$. In accordance with expectation, it has a weak (apparently logarithmic) dependence on $D_R$. For $D_R = 0.03$ one obtains $l = 0.10$. Note that at very small $D/J$ (large $R_f$) the finite size of the system begins to affect numerical results.  

\begin{figure}
\includegraphics[width=88mm]{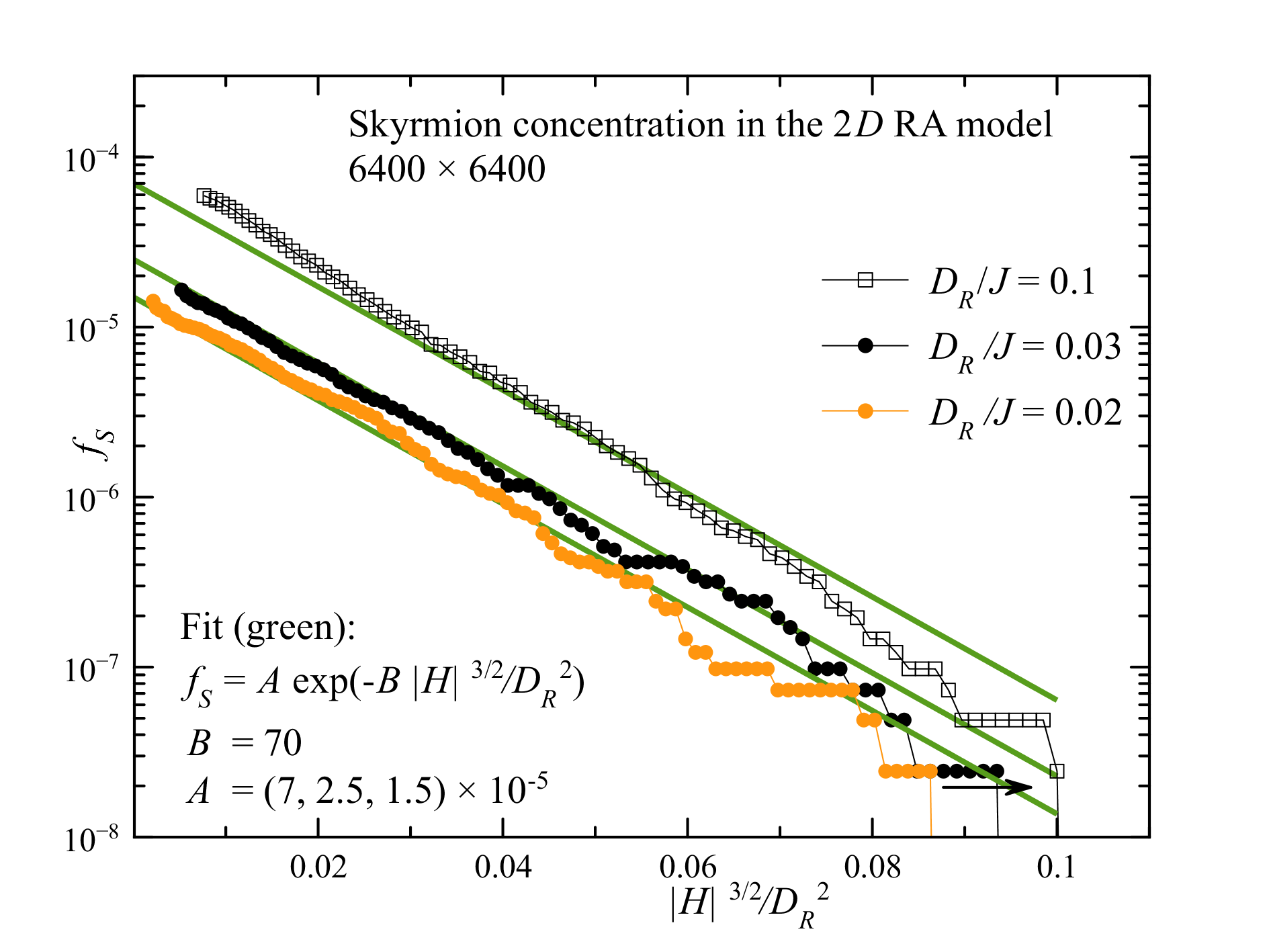}
\caption{Color online: Concentration of skyrmions as function of the magnetic field and RA strength.}
\label{concentration}
\end{figure}

Fig.\ \ref{concentration} shows concentration of skyrmions, $f_S$, as function  of $H$ and $D_R$. The expression in the fitting exponent can be written as $\left({|H|}/{H_c}\right)^{3/2} =64\pi\sqrt{2l}({|H|^{3/2}}/{D_R^2})$ in a remarkable agreement with the theory. Substituting here $l = 0.1$ one obtains $64\pi\sqrt{2l}\approx 89$, which is not far from the large factor in the fitting exponent in Fig.\ \ref{concentration}.  

In conclusion, we have demonstrated that a large concentration of skyrmions can be created in a magnetic film with quenched randomness. Good understanding of the field dependence of the concentration of skyrmions and their average size has been developed by analytical and numerical methods that agree with each other. Our findings should encourage experiments on skyrmions in disordered films.

This work has been supported by the grant No. DE-FG02-93ER45487 funded by the U.S. Department of Energy, Office of Science.

\end{document}